\documentclass[aps,prl,twocolumn,superscriptaddress]{revtex4-1}

\usepackage{graphicx}
\usepackage{amssymb}
\usepackage{amsmath}
\usepackage{color,soul}

\begin{document}

\title{Proposal for the Detection of Magnetic Monopoles in Spin Ice via Nanoscale Magnetometry}

\author{Franziska K.~K.~Kirschner}
\email{franziska.kirschner@physics.ox.ac.uk}
\affiliation{Department of Physics, University of Oxford, Clarendon Laboratory, Parks Road, Oxford, OX1 3PU, United Kingdom}

\author{Felix Flicker}
\altaffiliation{Present address: Rudolf Peierls Centre for Theoretical Physics, University of Oxford, Oxford OX1 3NP, United Kingdom}
  \affiliation{Department of Physics, University of California, Berkeley, California 94720, USA}
  
\author{Amir Yacoby}
  \affiliation{Department of Physics, Harvard University, Cambridge, Massachusetts 02138, USA}
\author{Norman Y.~Yao}
  \affiliation{Department of Physics, University of California, Berkeley, California 94720, USA}
  \affiliation{Materials Science Division, Lawrence Berkeley National Laboratory, Berkeley CA 94720, USA}
\author{Stephen J.~Blundell}
\email{stephen.blundell@physics.ox.ac.uk}
\affiliation{Department of Physics, University of Oxford, Clarendon Laboratory, Parks Road, Oxford, OX1 3PU, United Kingdom}

\date{\today}

\begin{abstract}
We present a proposal for applying nanoscale
magnetometry to the search for magnetic monopoles in the spin ice
materials holmium and dysprosium titanate.  Employing Monte Carlo
simulations of the dipolar spin ice model, we find that when cooled to
below $1.5\,$K these materials exhibit a sufficiently low monopole
density to enable the direct observation of magnetic fields from
individual monopoles. At these temperatures we demonstrate that
noise spectroscopy can capture the intrinsic fluctuations associated with
monopole dynamics, allowing one to isolate the qualitative effects
associated with both the Coulomb interaction between monopoles and
the topological constraints implied by Dirac strings.
We describe in detail three different nanoscale magnetometry platforms (muon
spin rotation, nitrogen vacancy defects, and nanoSQUID arrays) that can be
used to detect monopoles in these experiments, and
analyze the advantages of each. 

\end{abstract}

\maketitle
Although fundamental magnetic monopoles have so
far proven elusive, it has recently become
possible to study properties of monopole-like excitations
in condensed matter systems~\cite{kimballmilton}. The spin ices,
in particular dysprosium titanate (Dy$_2$Ti$_2$O$_7$, DTO) and holmium titanate
(Ho$_2$Ti$_2$O$_7$, HTO), have been identified as promising candidates
to host such elementary excitations~\cite{Harris1997,Castelnovo2007,Ryzhkin2005,Morris2009}.  The magnetic
rare earth metal
ions (Dy$^{3+}$ or Ho$^{3+}$) and the non-magnetic Ti$^{4+}$ ions are
arranged on two separate interpenetrating pyrochlore sublattices, each
consisting of a network of corner-sharing tetrahedra. The rare earth
moments ($\approx 10~\mu_{\rm B}$) are well-modeled as classical Ising spins, constrained by the crystal field
to lie along the local $\langle111\rangle$ axes. Exchange and dipolar
interactions acting in this lattice geometry result in the four spins in each
tetrahedron adopting a ground state in which two spins
point towards, and two away
from, the tetrahedron's center. This is termed the `ice rule', by analogy with protons in water
ice~\cite{Pauling1935,Harris1997}. This leads to a macroscopic degeneracy
in the ground state, with six possible spin configurations per
tetrahedron.

The elementary excitations in spin ice consist of single flipped
spins, which can fractionalize into a pair of monopoles, each
traveling through the lattice by successive spin
flips~\cite{Castelnovo2007}.  These monopoles manifest themselves as
sinks and sources of magnetization, corresponding to tetrahedra in the
3-in-1-out or 1-in-3-out configurations~\cite{Huse2003, Isakov2004,
  Henley2005}.  Including the dipolar interactions between spins, the
inter-monopole interaction has the form of Coulomb's law, completing
the analogue to fundamental magnetic monopoles~\cite{Fennell2009,
  Henley2010, Castelnovo2007} and allowing the spin ice state to be
described as a $U(1)$ classical spin liquid~\cite{Hermele2004}.
Unambiguous observation of the individual monopoles in spin ice would
not only confirm this theoretical picture, but would allow these
excitations and their dynamics to be studied directly.  Previous
attempts to identify the spin ice state share the common feature that
they infer monopole behavior from the monopoles acting {\it en masse}
\cite{Castelnovo2007,Morris2009,Fennell2009,Pomaranski2013}. The
direct measurement of the microscopic magnetic field from individual
monopoles remains an open challenge.

Recent developments have opened the door to new possibilities for the
detection and characterization of magnetic textures on the nanometer
scale.  In this Letter, we employ Monte Carlo numerical modeling to
provide both quantitative and qualitative predictions for what this
next generation of nanoscale magnetometers will be able to probe when
applied to spin ice at low temperatures.  We conclude that 
measurements of the
magnetic noise spectral density $S(\omega)$ will contain features
arising from
monopole dynamics due to both topological constraints arising from
Dirac strings, and long-range forces.
We propose and analyze three different detection platforms: muon spin rotation ($\mu$SR),  nitrogen-vacancy (NV)
magnetometry, and nanoscale arrays of superconducting quantum
interference devices (nanoSQUIDs).
By utilizing these techniques, it should be possible to experimentally constrain parameters such as the monopole density and hop rate as a function of temperature.

\begin{figure}[t]
  \centering
  \includegraphics[width=0.48\textwidth]{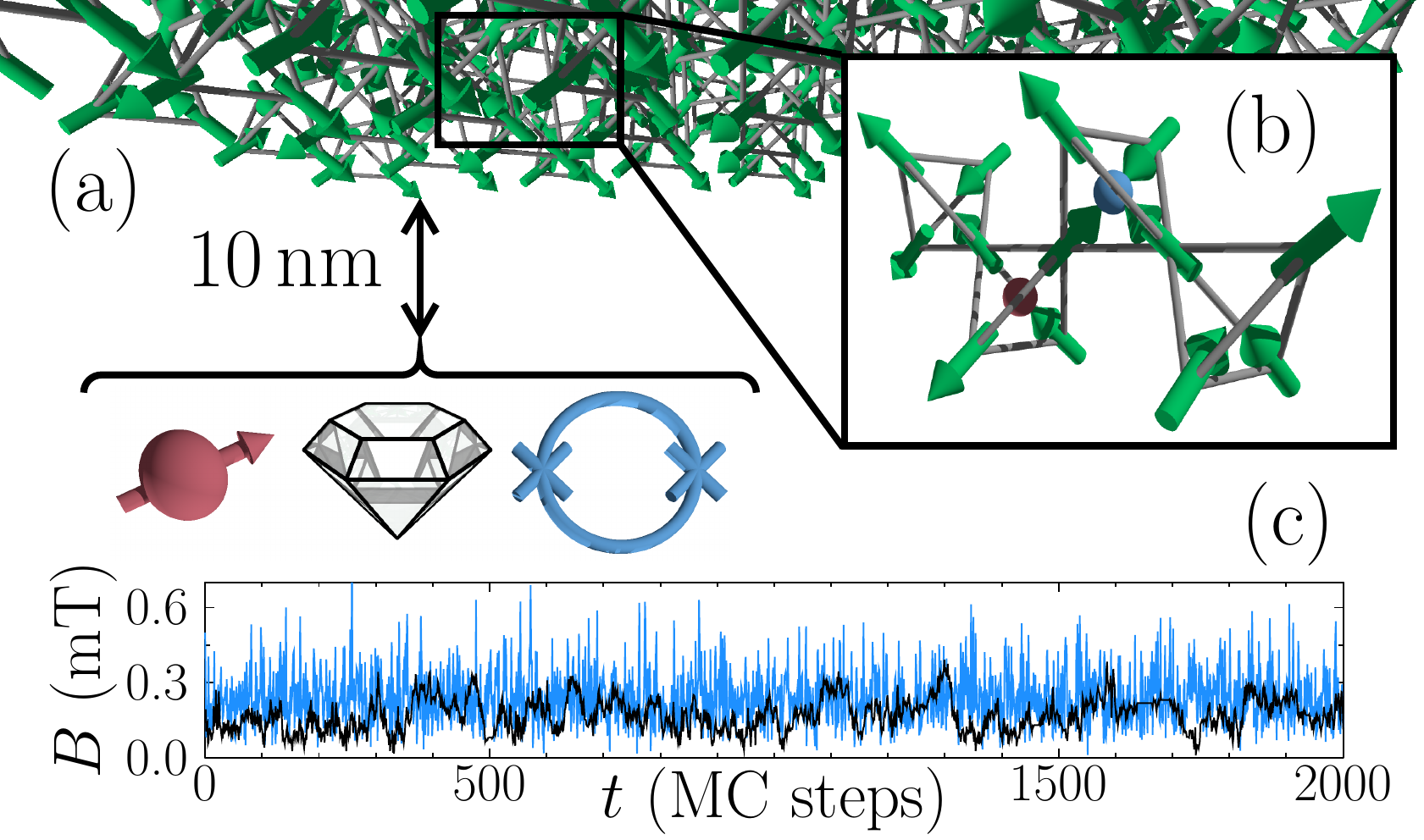}
\caption{(a) A schematic of the experimental arrangement for the techniques mentioned in this Letter, with a muon (left), NV centre in diamond (centre), and nanoSQUID (right) acting as ultrasensative magnetic probes $10\,$nm from the spin ice. (b) Four adjacent tetrahedra of Dy$^{3+}$ spins, showing the creation of a monopole-antimonopole pair. (c) The field fluctuations 10\,nm from the surface in DSIM at 4\,K (blue) and 1\,K (black).}
\label{fig1}
\end{figure}

\emph{Model}-- In order to make quantitative predictions we employ the full dipolar spin ice Hamiltonian:
\begin{equation}\label{Hamiltonian}
H\hspace{-0.2em} =\hspace{-0.2em} -J\hspace{-1.5em} \sum_{\langle \left( i , a \right) , \left( j , b \right) \rangle} \hspace{-1.2em}\mathbf{S}_i^a \cdot \mathbf{S}_j^b + D r_{\rm nn}^3 \sum_{\substack{i > j \\ a,b}} \frac{ \mathbf{S}_i^a \cdot \mathbf{S}_j^b }{ | \mathbf{R}^{ab}_{ij} |^3} - \frac{ 3 \left( \mathbf{S}_i^a \cdot \mathbf{R}^{ab}_{ij} \right) \left( \mathbf{S}_j^b \cdot \mathbf{R}^{ab}_{ij} \right) }{ | \mathbf{R}^{ab}_{ij} |^5 },
\end{equation}
for spin vectors $\mathbf{S}_i^a =\sigma_i^a \hat{\bf z}^a$, where
$\sigma_i^a = \pm 1$ and $\hat{\bf z}^a$ is the local Ising axis at the
tetrahedral sublattice site $\mathbf{r}^a$ for FCC lattice site
$\mathbf{R}_i$. The vector connecting two spins $\mathbf{S}_i^a$ and
$\mathbf{S}_j^b$ is thus given by $ \mathbf{R}^{ab}_{ij} =
\mathbf{R}_{ij} + \mathbf{r}^{ab}$. The exchange energy is $J \approx
-3.72~\rm{K}$ for DTO and $\approx -1.56~\rm{K}$
for HTO~\cite{DenHertog2000,Bramwell2001a}. The dipolar energy $D \approx
1.41~\rm{K}$ for both DTO and HTO~\cite{DenHertog2000,Bramwell2001a}. The first term in this Hamiltonian corresponds to
nearest-neighbor exchange interactions, with a nearest-neighbor
exchange coupling of $J_{\rm nn} = J/3$ (as the relative orientations
of nearest-neighbor $\langle 111 \rangle$ axes give $\hat{\bf z}^a \cdot
\hat{\bf z}^b\!=\!-1/3$). The second term corresponds to long-range dipolar
interactions. Though spin ices have been
predicted to undergo a first-order phase transition to long-range
order below $\sim 0.2$~K~\cite{Melko2001}, the equilibration time
rises dramatically on cooling and such order has
not yet been experimentally detected~\cite{Matsuhira2011,Pomaranski2013,HeneliusEA16}.

Simulations were carried out on $4 \times 4 \times 4$ unit cells of
spin ice using standard Monte Carlo (MC) procedures~\cite{Metropolis1953},
consisting of $10^4$ cooling steps and $5\times 10^3$ steps to measure the
stray magnetic fields of the system at temperatures between $4.5\,$K and
$0.5\,$K.
Periodic boundary conditions were used in the $\hat{\bf x}$ and
$\hat{\bf y}$ directions. The probe point for the
stray fields is placed $10~\rm{nm}$ from the sample in $\hat{\bf z}$, and
the time-dependence of the magnetic field $B(t)$ is calculated by
summing the dipolar field produced by all the spins in the film (a schematic of this experimental arrangement is shown in
Fig.~\ref{fig1}(a)).
MC simulations are often used determine thermal averages.
Here their use is restricted to creating a sample spin configuration at
a given temperature $T$, then to modeling the spin flip dynamics (and therefore the monopole motion). This approximation
has previously been suggested by experimental results~\cite{Bovo2013}.

Three different spin arrangements were studied in order to isolate the
characteristics of $S(\omega)$ arising from the individual contributions
to the Hamiltonian in Eq.~\ref{Hamiltonian}. This leads us to define three distinct models. (i) The dipolar spin ice
model (DSIM) uses the $J$ and $D$ parameters of DTO. (ii) The
nearest-neighbor spin ice model (NNSIM) has no long-range
interactions ($D=0$), but retains the topological constraint from
Dirac strings connecting monopoles to antimonopoles. $J$ is chosen so
that the monopole density is as for the first case, and $J>0$ 
ensures that the 2-in-2-out ground state is favored. (3) The all-in-all-out model (AIAO) has $D=0$ and
$J<0$. The ground state consists of tetrahedra with either four-in or four-out spin configurations. Thermal spin flips in this case are not monopole-like excitations, thus providing a control case.

\begin{figure}[t]
\centering
 \includegraphics[width=0.48\textwidth]{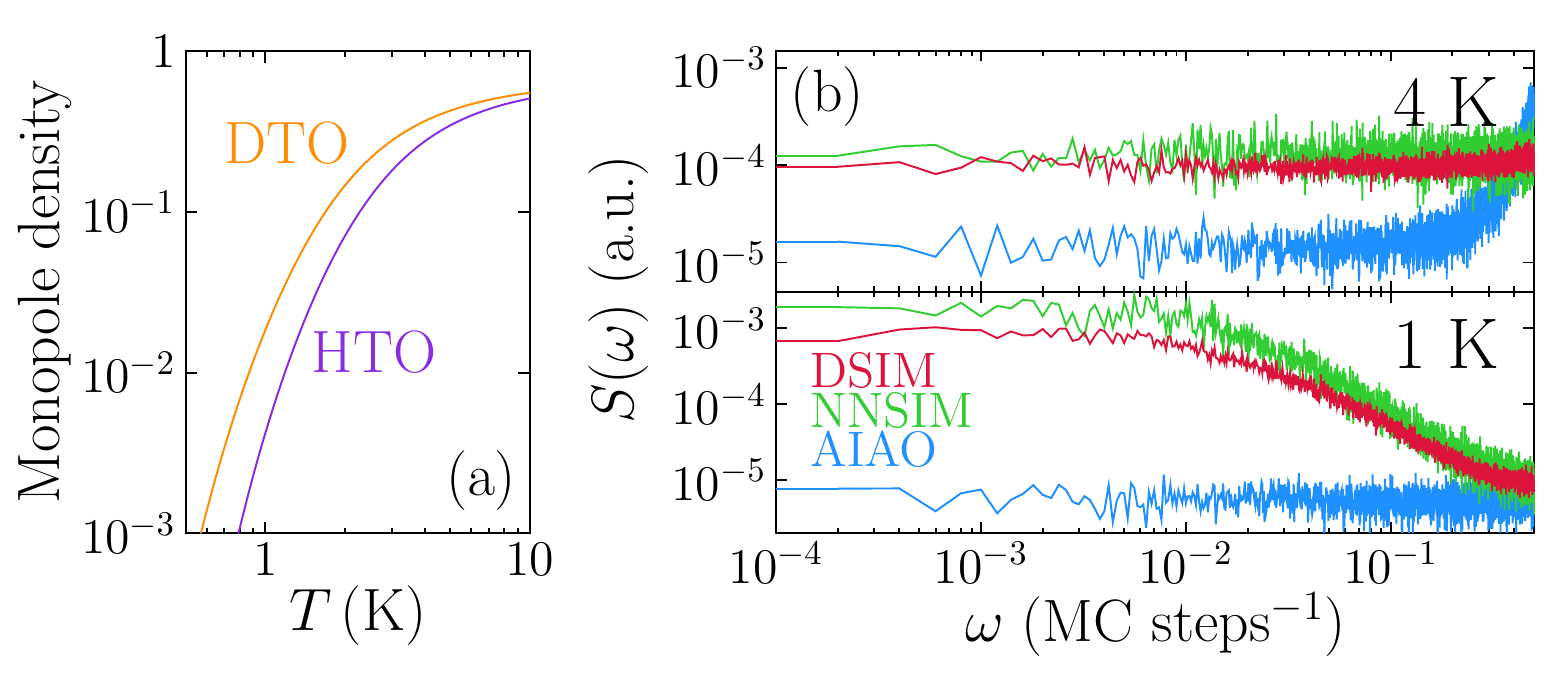}
\caption{ (a) Evolution of the monopole density in DTO and HTO calculated using Debye-H\"{u}ckel theory. (b) Frequency dependence of noise spectra $S(\omega)$ for DSIM, NNSIM, and AIAO [see main text for descriptions of the models] at 4\,K and 1\,K.}
\label{fig2}
\end{figure}

\emph{Results}-- In Figure~\ref{fig1}(c) we show the DSIM prediction
for the magnetic field measured at $4\,$K and $1\,$K.  At $4\,$K most tetrahedra are not in
the 2-in 2-out state and a rapidly fluctuating signal is observed.  At
1\,K the system is close to the ground state with a density of
monopoles per tetrahedron of $\approx 0.03$, as predicted by
Debye-H\"{u}ckel theory~\cite{Castelnovo2011}. The monopole density, calculated analytically using the methods described in Ref.~\cite{Castelnovo2011}, is slightly lower in HTO, shown in Fig.~\ref{fig2}(a).  At this density,
most tetrahedra obey the ice rules, but there are sufficiently many monopoles that some may
hop across the sample without annihilating, resulting in telegraph noise. By comparison, the AIAO
features no qualitative distinction between its high- and
low-temperature regimes. The amplitude of the field fluctuations are of order $0.1\,$mT,
well within the experimentally-measurable range of the aforementioned techniques.

The timescales in Fig.~\ref{fig1}(c) are in units of MC steps,
the shortest possible timescale on which a spin flip may occur in the
model. There has been considerable debate surrounding the timescale on
which spins flip in the physical systems at low temperature, with ac
susceptibility measurements suggesting a scale of $\sim 1~\rm{ms}$~\cite{Snyder2004,Jaubert2009,Quilliam2011,Kaiser2015,Tomasello2015}, and $\mu$SR measurements detecting spin dynamics on
a scale of $\sim 1~\mu\rm{s}$~\cite{Lago2007}. It is possible that this is
dependent on the relevant experimental time window ($10^0
- 10^{-4}~\rm{s}$ for ac susceptibility and $10^{-5} -
10^{-11}~\rm{s}$ for $\mu$SR). By performing the nanoscale
magnetometry measurements and comparing to the MC time step, the hop
rate can be deduced.

In addition to directly probing field fluctuations, complementary information is gained by measuring the noise spectral density $S(\omega)$, given as the Fourier transform of the autocorrelation function $A(\tau) = \int B(t)B(t+\tau)\,{\rm d}t$ \footnote{See Supplemental Material at [URL will be inserted by publisher] for phenomenological forms of the noise spectrum and additional simulations}.  In particular, the latter provides an ability to tune the frequency filter in order to isolate certain dynamical time-scales and to compare the amplitude of noise at these time-scales. As shown in
Fig.~\ref{fig2}(b), there is little low-frequency structure in
$S(\omega)$ at 4~K, corresponding to very little correlation between
the field measurements at different times. Fluctuations are predominantly
at the Nyqvist frequency [$\omega_{\rm Ny}=(2\times \mbox{MC time
    step})^{-1}$]. The lack of structure is evident in both DSIM and
NNSIM, and shows that it would be difficult to discern monopole
dynamics within the temperature range of $^4$He cryostats. For AIAO, a peak forms at
$\omega_{\rm Ny}$ as a result of rapid thermal fluctuations. The power
spectrum at 1~K, however, shows a clear difference between the DSIM,
NNSIM, and AIAO cases. The low-frequency structure seen in
Fig.~\ref{fig1}(c) is manifest in $S(\omega)$. It should be noted that
the frequency at which this structure occurs is on the order of
$\approx 10^{-2}$~MC steps$^{-1}$. The DSIM and NNSIM systems both
display low-frequency plateaus and a high-frequency power law, which
is absent in AIAO. This is a clear indication that the long-range
forces play a relatively small role in the monopole dynamics and
instead contribute more to the lowering of the overall stray field, as
evidenced in a smaller area under the $S(\omega)$ curve. This is
consistent with other theoretical studies, which suggest that samples
are largely dominated by single monopoles as opposed to closely bound
pairs~\cite{Castelnovo2011}. In AIAO, the field fluctuations are
smaller and contain no structure on longer timescales, which is
expected as it is always energetically favorable for spin flips to
undo shortly after creation.  We can make quantitative predictions of
the functional form of $S(\omega)$ by fitting power laws ($S(\omega) =
a \omega^b$) to the low- and high-$\omega$ regimes (see also \footnotemark[\value{footnote}] for power-law fits of a simpler toy model). DSIM, NNSIM, and
AIAO all display low-$\omega$ plateaus, with $|b| \sim
10^{-2}$. However, in the high-$\omega$ regime there are pronounced
differences between the three systems. For AIAO, $b>0$ at both low and
high temperatures ($b_{\rm 1\,K}^{\rm AIAO}=0.14(5)$, $b_{\rm 4\,K}^{\rm
  AIAO}=4.94(14)$), indicative of a system dominated by thermal spin
flips. Both DSIM and NNSIM have $b<0$ at low temperature
($b_{\rm 1\,K}^{\rm DSIM}=-1.29(1)$, $b_{\rm 1\,K}^{\rm NNSIM}=-1.56(3)$) which
increases at high temperatures ($b_{\rm 4\,K}^{\rm DSIM}=-0.09(2)$, $b_{\rm 4\,K}^{\rm NNSIM}=-0.08(4)$) as thermal spin
flips take over and monopole dynamics are no longer observable.

\emph{Experimental techniques}-- We now consider a number of different nanoscale magnetometry platforms which are promising candidates for detecting monopole behavior in spin ice materials. In addition to providing an analysis of the various regimes of operation, we highlight the complementarity of these measurements. 
The first experimental technique,
$\mu$SR, is usually a bulk probe, but a low-energy variant~\cite{morenzoni2004}
in which the energy of the muon beam can be continuously varied from
$0.5$ to $30\,$keV provides an extension of the technique which allows
depth-dependent studies of thin films and multilayered structures in
the range from $\sim$1  to $\sim200\,$nm.
This allows for experiments involving proximal magnetometry
\cite{salman2012}
in which the field close to an ultrathin magnetic layer can be probed.
$\mu$SR probes fields fluctuating on a timescale $\sim
10^{-11}$--$10^{-5}$\,s.
In the cases of zero-field $\mu$SR (in which
the polarization of the muons is measured in the absence of any
external magnetic fields) or longitudinal-field $\mu$SR (in which an
external field $B_{\rm L}$ is applied along the muon polarization direction),
the relaxation rate $\lambda$ of the muon polarization spectrum can be related
to the autocorrelation function of the components of ${\bf B}^{\perp}(t)$, the local field
transverse to the muon, using
%
$  \lambda = {\gamma_{\mu}^2 \over 2} \int_0^\infty \langle {\bf
    B}^\perp(t)\cdot{\bf B}^\perp(0) \rangle {\rm e}^{{\rm
      i}\gamma_{\mu}B_{\rm L}t}\,{\rm d}t$,
%
where $\gamma_\mu$ is the gyromagnetic ratio of the muon.
The relaxation rate $\lambda$ is then proportional to the power spectrum at
$\omega=\gamma_{\mu}B_{\rm L}$ and so in zero-field measures the
zero-frequency part of the power spectrum.
Inside spin ice the field at the muon site is very large, $\sim 1$\,T, but outside the sample the field is dominated by
a long-range stray field~\cite{Lago2007,Blundell2012}.  Our interest here is in
the nanoscale field close to the surface at low monopole
concentration. We note that the related technique of $\beta$-NMR, in
which
low energy ion implantation of hyperpolarized radioactive magnetic
resonance probes are employed instead of muons, can also be
used in this context~\cite{Xu2008}. A potential disadvantage of these
techniques is that the stopping profile of slow muons or other
polarized probes is not sharp, so that implantation occurs at a range
of depths.

The second method we consider is the use of single spin magnetometery
based upon Nitrogen-Vacancy (NV) point defects in diamond
\cite{Rondin2013, Lovchinsky16}. Each NV center constitutes an $S=1$ electronic
spin orientated along one of the four diamond
carbon-carbon bond directions. In addition to coherent manipulations
via resonant microwave pulses, the NV center's spin state can be
optically initialized and detected
\cite{Childress06, Maze2008, Shi2015}. We envision two possible setups
for NV-based monopole magnetometry. The first is a scanning NV
magnetometer, consisting of a diamond nanopillar attached to an AFM
tip~\cite{amir}. The second entails the placement of the spin ice
material in direct proximity to a bulk diamond surface containing a
shallow layer of NV centers approximately $\sim 5-10\,$nm deep.
To detect the characteristic signatures of individual magnetic monopoles, the NV can be utilized in two operational modes: 1) direct measurement of the stray magnetic field from a monopole and 2) spectroscopy of the ac magnetic noise generated by the dynamics and fluctuations of a dilute monopole density. 

In the case of dc magnetometry, one would observe Zeeman shifts in the NV resonance frequency (\emph{e.g.} between the $|m_s = 0\rangle$ and $|m_s = -1\rangle$ spin states) in real time in order to measure the stray field of a monopole as it passes through the sensing volume of a single shallow NV center (Fig.~\ref{fig1}).
This can be achieved using either Ramsey spectroscopy or continuous-wave optically detected magnetic resonance spectroscopy. The field sensitivity of this approach is limited by $T^*_2$, the NV dephasing time, leading to a sensitivity $\sim$5\,$\mu$T/$\sqrt{\textrm{Hz}}$ \cite{Maletinsky2012}.  Assuming an integration time
 $\sim 250\mu$s this enables a field sensitivity of approximately
 $\sim 0.3$~mT and a corresponding dc sensing volume of approximately
 $\sim 10$~nm surrounding the NV center. This sensitivity is sufficient to detect the real time dynamics of individual monopoles (see Fig.~\ref{fig1}).
 
For ac magnetometry, we are interested in the individual Fourier components of the time-varying magnetic field generated by the dynamics of a low density of magnetic monopoles. To generate such a frequency filter, the NV center is manipulated using a series of periodic microwave pulses separated by a free-evolution time $\tau$. This modulation creates a narrow band-pass frequency filter at $1/\tau$; by varying the free-evolution time, one can map out the noise spectral density associated with magnetic fluctuations. Compared to the DC case, the key advantage of this approach is that the field sensitivity is no longer limited by $T^*_2$ , but rather by $T_2$, the intrinsic spin decoherence time of the NV center. For shallow-implanted NVs, this yields a field sensitivity $\sim 50$~nT/$\sqrt{\textrm{Hz}}$ \cite{Maletinsky2012}. Assuming an integration time $\approx 250\,\mu\rm{s}$ this enables a field sensitivity of approximately $\approx 3\,\mu\rm{T}$ which is well within the desired sensitivity.

The third experimental method employs nanoSQUIDs.  The superconducting
quantum interference device (SQUID) is an extremely sensitive detector
of magnetic flux, and decreasing its size leads to a highly versatile
sensor of local magnetic fields~\cite{wernsdorfer2009,Granata2016},
with a spin sensitivity that has reached below a single Bohr magneton
\cite{vasyukov}. 

The experimental techniques considered in this Letter have different
advantages and drawbacks. The nanoSQUID technique, despite excellent
sensitivity and the ability to work at the required temperature
regime, may suffer from a large spatial averaging of fields across the
area of the sensor ($\sim 10^5$~nm$^2$).  This is much larger than
that
for the polarized probes and NV
centers for which the active sensor is essentially point-like.
Low energy muons, though point
probes, can only be implanted over a range of depths.  Thus a
suitable overlayer can be deposited on the surface of spin ice and
muons implanted into it, but the observed signal will average over a
spread of distances between the spin ice surface and the probe,
although the mean distance can be varied by varying the implantation
energy.  NV centers may be much better in this regard, though
measurements need to be obtained with an applied magnetic field, which
is not the case for $\mu$SR.

At the time of writing, both $\mu$SR and NV-magnetometry experiments
are limited to temperatures above 4.2~K by the use of $^4$He
cryostats. When such experiments are able to reach temperatures of
$\approx$1.5~K, there are clearly-discernible, qualitative signatures
of two key characteristics of magnetic monopoles: their Coulomb
interaction and topological constraints deriving from Dirac
strings. Unconstrained thermal spin flips feature a Debye-type noise
spectral density, matching that of Brownian motion, with a
low-frequency plateau followed by a turnover to inverse-square flicker
noise at characteristic timescale $\tau$. Coulomb interactions between
Brownian particles decrease the timescale $\tau$ through recombination
of particle/antiparticle pairs. In general, one expects a range of
timescales to be important in $S(\omega)$ originating from a slowing
of dynamics heading out of equilibrium at the lowest temperatures.
The nanoscale probes proposed in this letter can cover a wide range of
frequencies from $10^0-10^{-4}$\,s (NV magnetometers) through
$10^{-5}-10^{-11}$\,s ($\mu$SR) in order to probe the relevant
dynamics and to understand the scaling between experimental timescales
and those used in MC simulations, both of which can be tuned by
temperature.

Two further modes of operation which can be used in the identification
and characterization of monopoles with these probes are as follows.
We note the possibility of making two-point correlation measurements
using two probes in NV-magnetometry.
Such measurements could be used to time the motion of single monopoles, helping
constrain system parameters such as the magnetic charge when used in
combination with applied $B$ fields. A further possibility is to use
one magnetized tip and one measurement tip to probe the response to
local perturbations.

In summary, we have presented a roadmap for future experiments of
monopole behavior using nanoscale magnetic probes of the noise
spectrum of the dipolar field measured very close to the surface of
spin ice.  The techniques have varying advantages and disadvantages,
and need to be extended to the 1~K regime, where the monopole density
is sufficiently low that observation of individual magnetic
monopoles is possible. Such an observation would open up an era of
direct measurement of monopole transport in these topologically-constrained systems and provide new insight into the classical $U(1)$
spin liquid.

\emph{Acknowledgments}: The authors would like to thank M.~J.~P.~Gingras and D.~Santamore for helpful discussions. We thank EPSRC (UK) for funding support.  F.~K.~K.~K. thanks Lincoln College Oxford for a doctoral studentship. F.~F. acknowledges support from a Lindemann Trust Fellowship of the English Speaking Union, and an Astor Junior Research Fellowship from New College, Oxford. N.Y. is supported by the LDRD Program of LBNL under US DOE Contract No. DE-AC02-05CH11231. A.Y. is supported by the Gordon and Betty Moore Foundation’s Emergent Phenomena in Quantum Systems (EPiQS) Initiative through grant GBMF4531. A.Y. is also partly supported by Army Research Office grant W911NF-17-1-0023.

\bibliographystyle{apsrev4-1}
\bibliography{references}
\

\end{document}